\documentclass[10pt]{article}

\pdfoutput=1
\usepackage[pdftex]{color}
\usepackage{amssymb}
\usepackage{amsthm} 
\usepackage{amsmath}
\usepackage{latexsym}
\usepackage{amscd}
\usepackage{graphicx}
\usepackage[pdftex, colorlinks=true, citecolor=green]{hyperref}
\usepackage{lscape}
\usepackage{setspace}
\usepackage{multirow}
\usepackage{wrapfig}

\newcommand{\ben}{\begin{equation}}     
\newcommand{\eeqn}{\end{equation}}
\newcommand{\bey}{\begin{eqnarray}}
\newcommand{\eey}{\end{eqnarray}}

\begin{document}

\begin{flushleft}
{\Large
\textbf{Gap junctions and synchronization clusters in TRN}
}

\emph{Anca R\v{a}dulescu$^1$, Michael Anderson$^1$}

\emph{Department of Mathematics, SUNY New Paltz, NY 12561}

\end{flushleft}

\begin{abstract}
    The Thalamic Reticular Nuclei (TRN) mediate processes like attentional modulation, sensory gating and sleep spindles. The GABAergic inter neurons in the TRN are know to exhibit widespread synchronized activity patterns. One known contribution to shaping synchronization and clustering patterns in the TRN is coming from the presence of gap junctions. These are organized in specific connectivity architectures, that have been identified empirically through dye and electrical coupling studies.
    Our study uses a computational model in conjunction to implement realistic connectivity schemes in a small network. We explored the potential effects of the size, strength and distribution of gap junctional clusters on the synchronization patterns in TRN, and how these effects are modulated by other factors, such as the level of background inhibition.
\end{abstract}

\section{Introduction}

\subsection{TRN rhythms and gap junction coupling}

The Thalamic Reticular Nuclei (TRN) mediate processes like attentional modulation, sensory gating and sleep spindles. The GABAergic neurons that comprise the TRN, which surround and inhibit the neurons in the thalamic relay nuclei~\cite{crabtree2018functional,contreras1997spatiotemporal} exhibit widespread synchronized activity patterns. It has been known for a few decades that these patterns depend on many factors, and are not locally restricted, extending broadly to thalamic and cortical regions~\cite{krosigk1993cellular,steriade1993thalamocortical}. When special circumstances are created, the TRN can also generate abnormal rhythms~\cite{slaght2002activity,mccormick2001cellular}, with deficit in TRN functions being tied to mental illnesses such as bipolar disorder and schizophrenia. Despite their potentially crucial importance to cognitive function and behavior, pattern formation and synchronization mechanisms within the TRN are not fully understood.

One known contribution to shaping synchronization and clustering patterns in the TRN is coming from the presence of gap junctions (GJ). Recent research has shown that about one third to one half of the neurons in the TRN of rodents form electrical synapses ~\cite{crabtree2018functional}. Gap junctions are frequent in circuits of inhibitory interneurons~\cite{connors2004electrical,simon2005gap}, are known to be involved in electrical signaling that modulates firing rates, and may lead to synchronization of spiking and subthreshold activity~\cite{vervaeke2010rapid,hjorth2009gap}.

GJ-coupled neurons in the TRN are structured in clusters, which have been identified empirically through dye~\cite{lee2014two} or electrical coupling studies~\cite{long2004small}. It was shown that there are two functionally distinct types of GJ-coupled neural clusters, defined by their spatial configuration as ``elongated'' and ``discoid''~\cite{lee2014two}. In addition, these two cluster types also differ in their projections to
thalamic somatosensory regions~\cite{lee2014two}. It is well known that the spatial organization of neuron clusters coupled by GJs is an important determinant of network function, yet it is poorly described for nearly all mammalian brain regions.   

Die coupling studies found that there are in average 9 cells, and up to 24 cells in any TRN gap-junction cluster~\cite{crabtree2018functional,lee2014two}. Because they contribute to coordination of inhibitory activity, it is to be expected that network architectures of inhibitory TRN neurons with different distributions of GJ coupled clusters can produce different phase coherence patterns in the target regions~\cite{tiesinga2008regulation}. Indeed, recent studies suggest that the power of GJs to coordinate neuronal activity is dependent on the spatial organization of coupled neuronal clusters, including their size and geometric patterns within the TRN and broader circuits. For example, it was shown that in the absence of chemical synaptic transmission, subthreshold rhythms and the
evoked action potentials were well synchronized between closely spaced, electrically coupled pairs; however, rhythms in noncoupled cells were not synchronized. This suggests that only in the presence of chamical synaptic background can GJ coordinate spindle-frequency rhythms among cell clusters in the TRN~\cite{long2004small}. Despite its seeming importance, the contribution of the spatial organization of GJ-coupled networks to the functional dynamics of TRN circuits is poorly understood in almost all parts of the brain. In this study, we use a computational model to explore the potential effects of the size, strength and distribution of gap junctional clusters on the synchronization patterns in TRN.

\subsection{Modeling of the TRN}

In a seminal paper from 1993, Golomb and Rinzel used a reduced Hodgkin-Huxley type model~\cite{golomb1993dynamics} to analyze the synchronization and clustering in the TRN~\cite{golomb1994clustering}. In this model, only chemical synapses were considered between TRN interneurons, in a rigid, all-to-all configuration. Even with this basic assumption, the authors showed that a network of 100 identical neurons, but with random initial conditions, can separate into different numbers of synchronized clusters (from 2 to 5), depending on the strength of the synaptic conductance between node pairs. The model will be further reviewed in the Methods section, since it is the foundation of this study. Departing from the original model, one possible direction of investigation could be that of considering more complex chemical synaptic architectures than an all-to-all scheme. 

A different direction, approached by Kopell and Ermentrout~\cite{kopell2004chemical} in 2004 (ten years after the original model), is to study the effect of overlaying gap junctions onto the all-to-all chemical connectivity structure. Using a combination of analytical and computational techniques, the study showed that the electrical and inhibitory coupling  play complementary
roles in the coherence of rhythms in inhibitory networks of interneurons (such as those in the TRN). More precisely, inhibition was efficient at equalizing the effects of random initial conditions, but had little effect on coherence. In contrast, electrical coupling was found to act in the direction of pulling together the voltages between spiking intervals, and of minimizing suppression during spiking intervals. Even a small amount of added electrical coupling was found to efficiently contribute to phase coherence when GABA inhibition was not sufficiently strong to accomplish the result in a heterogeneous network. More precisely, when electrical synapses with conductance 10-20\% of the synaptic conductance were added, coherence was sharply increased. To show that this was not just a consequence of the all–all connectivity, the experiment was also performed with cells arranged in a line, with a 10 or 20 nearest neighbors connectivity scheme, producing similar results. 

Based on newer data from dye studies and electrode recordings, finer information is available on the particular geometry of the gap junction distribution in the case of the TRN. To the best of our knowledge, these aspects have not yet been taken into consideration in any model to date. In our study, we revisit the the original Golomb-Rinzel model, with the goal of investigating the complementary roles of chemical and electrical synapses in the context of realistic connectivity schemes.

\section{Methods}

We build upon the traditional model proposed by Golomb and Rinzel~\cite{golomb1994clustering} for synchronization and clustering in TRN. This is a reduced Hodgkin-Huxley type system of $3N$ equations, describing membrane dynamics under all-to-all chemical synaptic interactions within a population of $N$ inhibitory neurons:

\begin{eqnarray}
\label{mother_eq}
C\frac{dV_i}{dt} &=& I_\text{Ca}(V_i,h_i)+I_\text{L}(V_i) - \frac{g_\text{syn}}{N} (V_i-V_\text{syn}) \sum_{i=1}^N s_i(t)\\
\frac{dh_i}{dt} &=& k_h(V_i)[h_\infty(V_i) - h_i]\\
\frac{ds_i}{dt} &=& k_f \cdot s_\infty(V_i)(1-s_i) - k_rs_i
\end{eqnarray}

\noindent where $V_i$, $h_i$ and $s_i$ represent the membrane potential, the inactivation variable of the calcium channel and the synaptic variable respectively, for the $i$th out of the $N$ neurons in the model population. The fast activation variable of the calcium channel is considered to depend instantaneously on the voltage as $m_\infty(V_i)$ (defined below). The currents in the system are defined as

\begin{eqnarray}
I_\text{Ca}(V,h) &=& -g_\text{Ca} m_\infty^3(V)h(V-V_\text{Ca}) \text{ and }\\ \nonumber \\
I_L(V) = &=& -g_L(V-V_L)
\end{eqnarray}

\noindent and

\begin{eqnarray}
k_h(V) = \frac{\phi \exp[-(V-\theta_{ht})/\sigma_{ht}]}{h_\infty(V)}
\end{eqnarray}

\noindent with

\begin{eqnarray}
m_\infty(V) = {\cal S}_{\theta_m,\sigma_m}(V), \quad h_\infty(V) = {\cal S}_{\theta_h,\sigma_h}(V), \quad s_\infty(V) = {\cal S}_{\theta_s,\sigma_s}(V)
\end{eqnarray}

\noindent all defined by the sigmoidal dependence 
\begin{eqnarray}
{\cal S}_{\theta,\sigma}(V) = \frac{1}{1+\exp[-(V-\theta)/\sigma]}
\end{eqnarray}

\noindent with corresponding values of the parameters $\theta$ and $\sigma$. In the original Rinzel model, the system parameters are chosen to represent biophysically the reticular thalamic nucleus (RTN), as follows: $C$ = $10^{-3}$ F/cm$^2$,  $g_\text{Ca}$ = 0.5 mS/cm$^2$, $g_L$ = 0.05 mS/cm$^2$, $V_\text{Ca}$ = 120 mV, $V_L$ = -60 mV, $V_\text{syn}$ = -80 mV, $k_f$ = $10^3$ 1/s, $\phi$ = $2 \cdot 10^3$ 1/s, $\theta_m$ = -65 mV, $\sigma_m$ = 7.8 mV, $\theta_h$ = -81 mV, $\sigma_h$ = -11 mV, $\theta_s$ = -45 mV, $\sigma_s$ = 2 mV, $\theta_{ht}$ = -162.3 mV, $\sigma_{ht}$ = 17.8 mV, $k_r$ = 50 1/s. The synaptic conductance $g_{syn}$ is a key parameter in our study, as explained below.

\begin{figure}[h!]
\begin{center}
\includegraphics[width=0.4\textwidth]{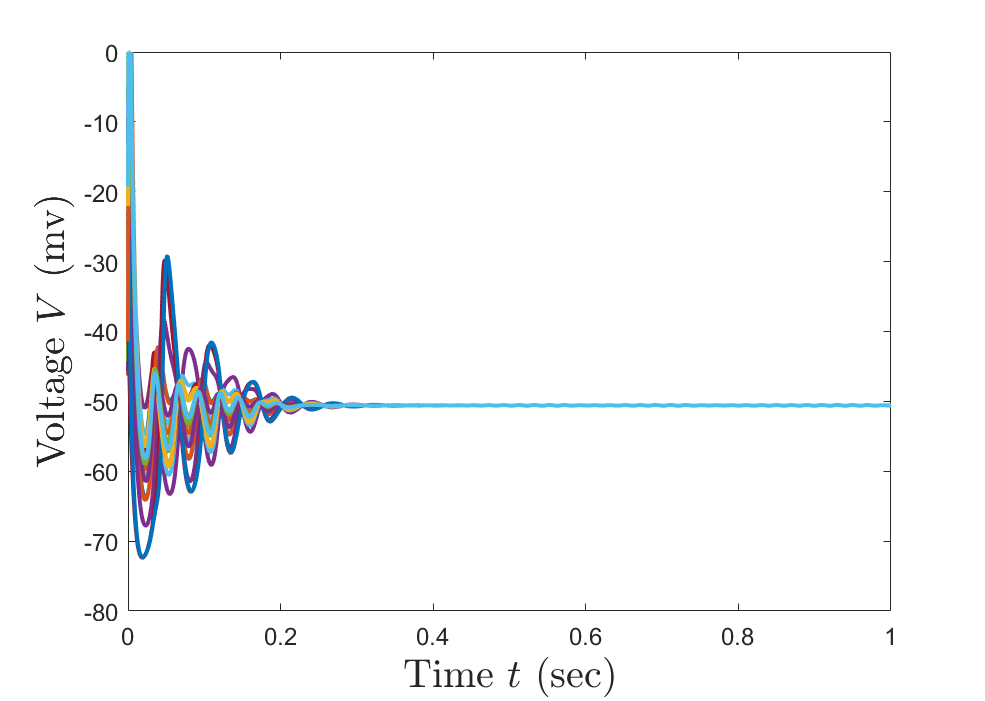}
\quad \quad
\includegraphics[width=0.4\textwidth]{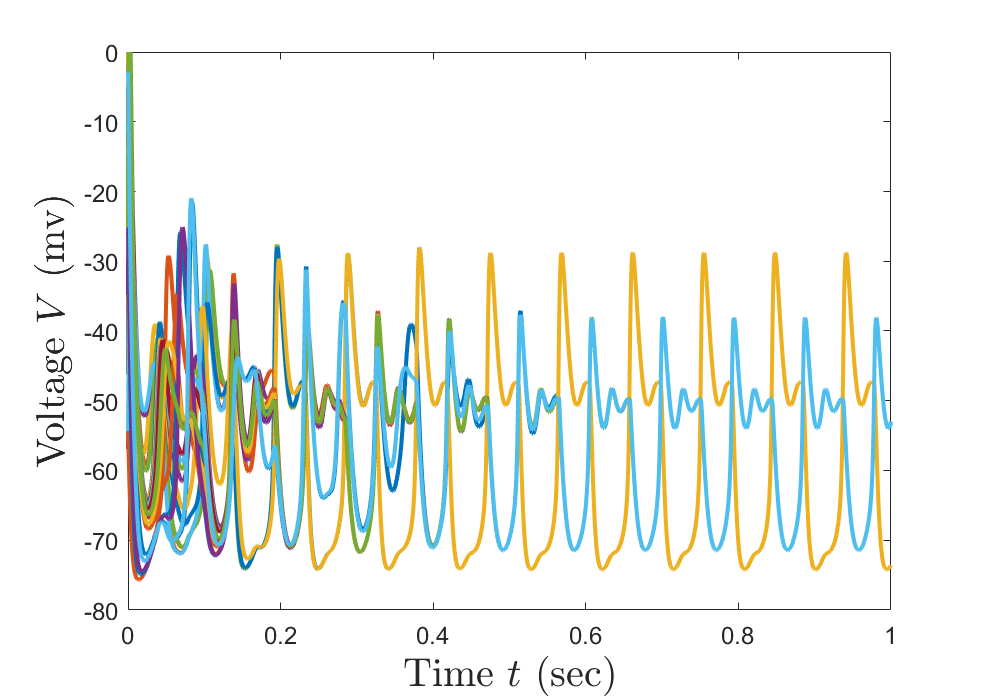}\\
\includegraphics[width=0.4\textwidth]{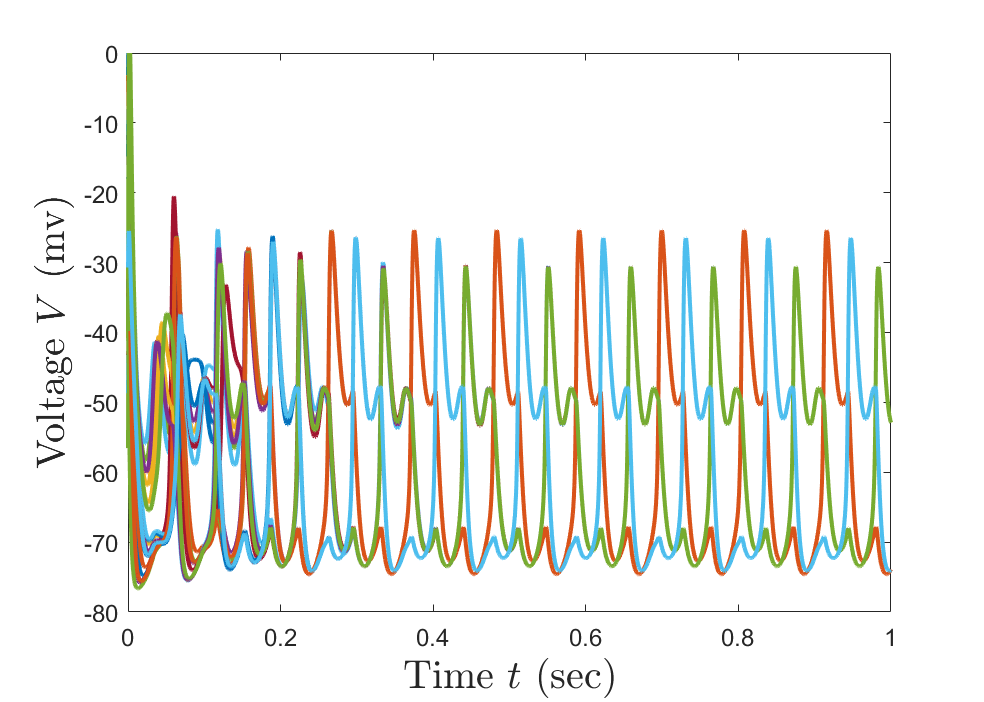}
\quad \quad
\includegraphics[width=0.4\textwidth]{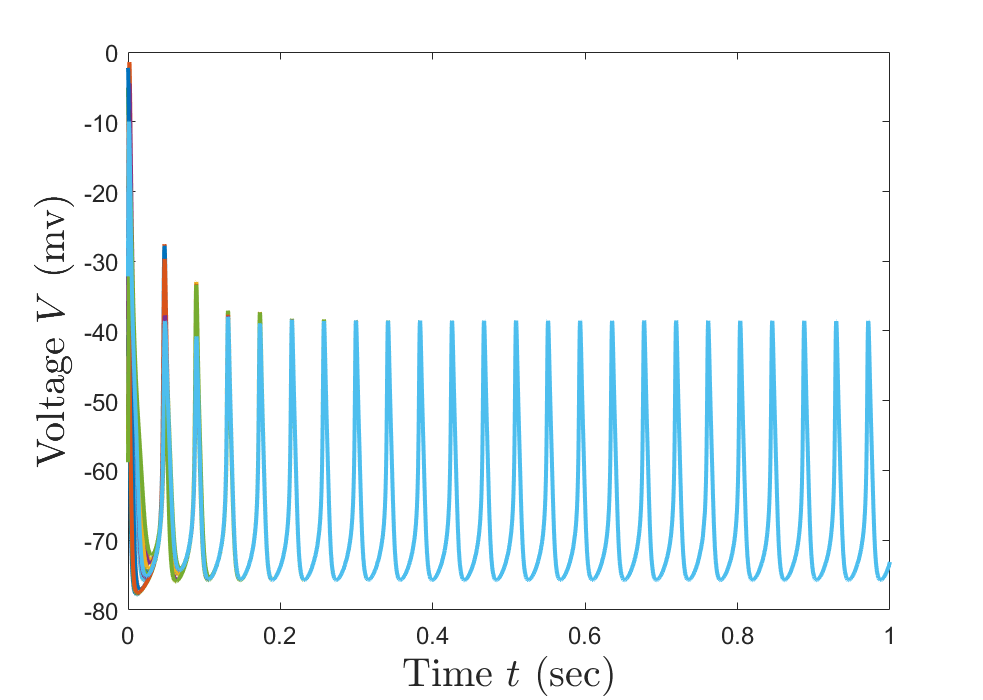}
\end{center}
\vspace{-5mm}
\caption{\emph{\small {\bf Examples of clustering behavior in the Golomb and Rinzel model.} {\bf A.} Damp oscillations to a stable equilibrium for $g_{syn}=0.2$; {\bf B.} convergence to a two-cluster oscillation for $g_{syn}=0.3$; {\bf C} convergence to a three-cluster oscillation for $g_{syn}=0.4$; {\bf D.} convergence to fully synchronized behavior for $g_{syn}=0.65$.}}
\label{Rinzel_solns}
\end{figure}

In the Golomb-Rinzel model, all neurons settle, possibly after a transient window, into clusters with temporally synchronized activity. In their work, the authors showed that, in the presence of all-to-all chemical synapses, the number, structure and duty cycle of the clusters can be controlled by only changing the synaptic conductance $g_{syn}$. Figure~\ref{Rinzel_solns} illustrates a few different clustering behaviors that the model exhibits in an all-to-all inhibitory TRN network with $N=20$ neurons, for different values of the conductance $g_{syn}$ and random initial conditions. For small values of $g_{syn}$ (e.g., $g_{syn}=0.2$), the neurons are not excitable, and the system quickly dampens the oscillations. When increasing $g_{syn}$, the oscillations become persistent, and will separate first into two out of phase synchronized clusters (e.g., for $g_{syn}=0.3$), then into three clusters (e.g., for $g_{syn}=0.4$). For large enough values of $g_{syn}$, the all cells synchronize. While the behavior of TRN GABA-ergic all-to-all networks, as reflected in the original model, has been extensively studied, our interest lies in understanding the contributions of the gap-junctional subnetworks that have been found in literature to overlay the synaptic inhibitory network. 

Starting with this basic system describing behavior in an the all-to-all inhibitory network, we add the gap junction connections between fixed subsets of neurons, specified by an adjacency matrix $A_{ij}$. With this modification, equation~\eqref{mother_eq} becomes:

\begin{eqnarray}
\label{mother_eq2}
C\frac{dV_i}{dt} &=& I_\text{Ca}(V_i,h_i)+I_\text{L}(V_i) - \frac{g_{syn}}{N} (V_i-V_\text{syn}) \sum_{j=1}^N s_i(t) - \frac{g_{el}}{M_i} \sum_{j=1}^N A_{ij}(V_i-V_j)
\end{eqnarray}

\noindent where $g_{el}$ is the gap junction conductance, and $M_i$ is the total number of gap junctions made on the $i$th neuron. According to the existing literature on gap junction architecture in the TRN, we assumed, throughout our simulations, that neurons connected by gap junctions are organized in disjoint, connected, all-to-all clusters of different sizes. To implement these conditions, we considered a circular organization of the nodes, so that if a gap junction connection exists with the $k$-th neighbor around the circle, then there are gap junctions with all the nearer neighbors $1$ to $(k-1)$.

\begin{figure}[h!]
\begin{center}
\includegraphics[width=\textwidth]{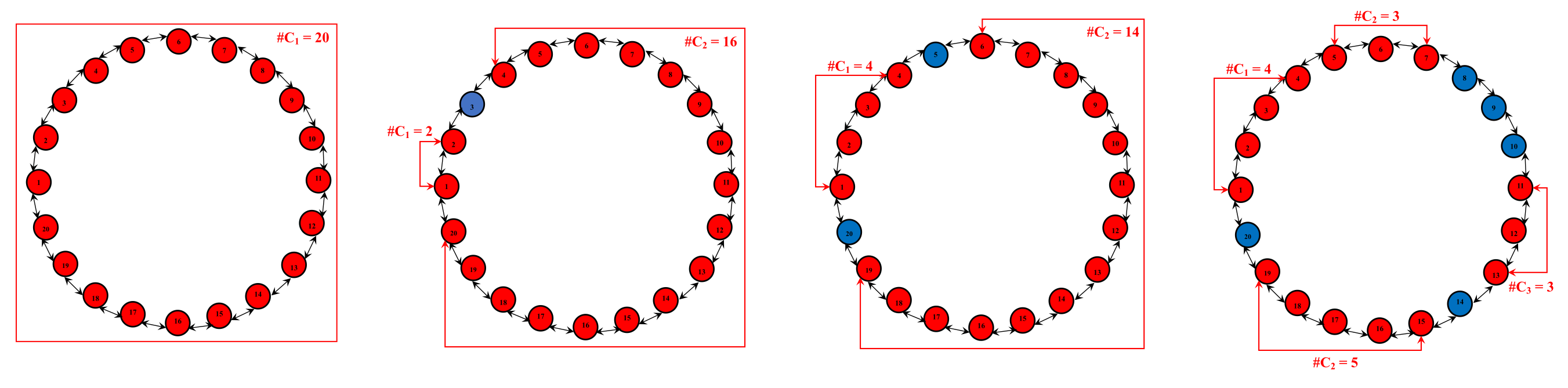}
\end{center}
\vspace{-5mm}
\caption{\emph{\small {\bf Diagrams illustrating the gap-cluster structures used in our simulations.} Each panel represents one network. All networks have $N=20$ neurons, organized in a circle, and labeled $1-20$. These neurons and are all connected to all by chemical synapses (represented only schematically by black arrows between all nearest neighbors,  to avoid crowding the figure).  The neurons involved in gap-clusters are shown in red, and the gap cluster structure is pointed out schematically by brackets, labeled with the cluster number and its size. The gap-junction adjacency matrices corresponding to all cases are shown in the Appendix.}}
\label{networks}
\end{figure}

Within this setup, we aim to study the dependence of the system's behavior on the values of the two conductance parameters $g_{syn}$ and $g_{el}$, and on the gap junction distribution (number, size and position of the structural clusters connected by gap junctions). As a measure of the system's dynamics, we compute the number of synchronization clusters formed asymptotically in the network (estimated based on simulations run for $T=500$ seconds, with integration step $h=0.01$ sec). As a matter of linguistics, we are working with two distinct concepts of clustering, both crucial to the study, and too well established in the literature to change the nomenclature entirely. To avoid confusion, we will use the term ``gap-clusters'' in relation with the geometric distribution of neuron subsets connected by gap junctions, and we will use the term ``synch-clusters'' to refer to groups of cell with temporally synchronized activity.

For our exploration, we considered a network of $N=20$ neurons, with conductances $g_{syn} \in [0.2,0.7]$ and $g_{el} \in [0,0.15]$. The parameter ranges were informed by our preliminary simulations, suggesting that the relevant behaviors occur within this locus. We considered four schemes for the gap-clusters structure, all within the empirical characteristics discussed in the introduction. The schemes are illustrated in Figure~\ref{networks}, ranging from two (one large, one small) gap-clusters to four small gap clusters, following the conditions described above. A larger network size would of course open access to more diverse configurations, but would also increase the computational cost. For our current study, we considered $N=20$ as a good compromise, optimizing between size and computational power. In this paper, we will be tracking primarily the number of synchronization clusters, and not as much their geometry, fraction of neurons involved in each cluster, or their distributions among clusters. While these are important questions to ask, they go beyond the scope of our paper. In all that follows, asymptotic attractors will be defined uniquely to transcend these issues, though the set of trajectories described by the nodes (i.e., states with identical synch-cluster trajectories, but with the nodes distributed differently between these synch-clusters, will be considered to represent the same attractor).

\section{Results}
\subsection{All-to-all gap junction connections}

To start fixing ideas, consider first a network of $N=20$ neurons which are all-to-all connected by both inhibitory synapses and gap junctions (the scenario in Figure~\ref{networks}a, with a unique gap cluster of size $C_1=20$ nodes). For this fixed underlying network, we want to study the effect of increasing the gap junction conductance $g_{el}$. In particular, we aim to test the theory, based on existing literature, that more spread-out and stronger gap junction transmission promotes synchronization within the system.

\begin{figure}[h!]
\begin{center}
\includegraphics[width=\textwidth]{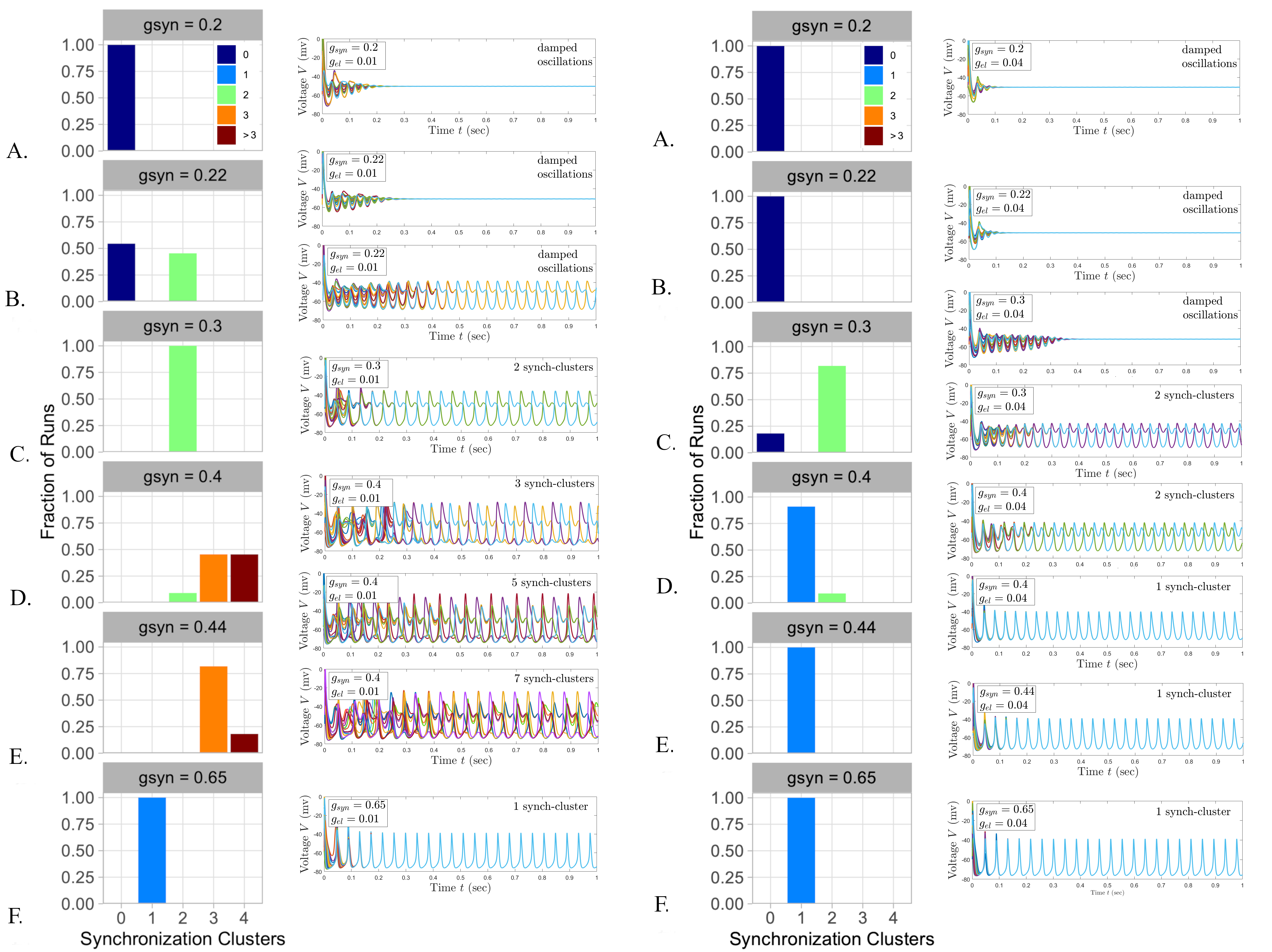}
\end{center}
\caption{{{\bf Changes in clustering behavior when $g_{syn}$ increases}, for fixed $g_{el}=0.01$ (panels A1-F1) and for $g_{el}=0.05$ (panels A2-F2). In each case, the left panels show the fraction of each behavior based on a sample of $K=11$ runs with random initial conditions, and the right panels illustrate examples of those behaviors in $(V,t)$ coordinate axes. From top to bottom, the value of $g_{syn}$ increases, as shown in the legends.}}
\label{horizontal_slices}
\end{figure}

Recall that, in absence of gap junctions (i.e., $g_{el}=0$), increasing the synaptic conductance $g_{syn}$ pushes the system through a succession of clustering patterns similar to those shown in Figure~\ref{Rinzel_solns}. Our simulations show that this changes when introducing even a small positive conductance $g_{el} = 0.01$, as shown in Figure~\ref{horizontal_slices}, panels A1-F1. Specifically, as $g_{syn}$ is increased, the system transition from damped oscillations (panel A1) to two synch-clusters (panel C1), through a transition region (around $g_{syn}=0.22$, panel B1) where these two attractors coexist, and can be both reached for different initial conditions. For higher synaptic strength ($g_{syn}$ around 0.37-0.45), the system gradually gains access to an attractor characterized by three or more synch-clusters (our simulations corresponding to panels D1  and E1 suggest up to nine distinct synch-clusters in this case, counted at the time $T=500$ secs when simulations were terminated). If $g_{syn}$ is increased beyond this range, the system gradually transitions into a fully synchronized regime (panel F1), through a transient bistability region where both adjacent attractor types are accessible. Overall, the emergence of stable oscillatory states with more than three clusters for small positive $g_{el}$ is surprising, and goes against the common belief that more gap junctions induce higher synchronization. 

Next, we perform a similar exploration to verify if this remains the case for stronger electrical conductances. The results of the analogous simulation for $g_{el}=0.05$ are illustrated in Figure~\ref{horizontal_slices}, panels A2-F2. They show damped oscillations for small $g_{syn}$ (panels A2 and B2), followed by a relatively sharp transition to a two synch-cluster state as $g_{syn}$ is increased (C2 and D2), then by a quick transition to full synchronization (panels E2 and F2). The simulation suggests that, for this larger value of $g_{el}$, the system no longer has access to  states characterized by three or more synch clusters, which is in agreement with the idea of gap junctions promoting synchronization.

\begin{figure}[h!]
\begin{center}
\includegraphics[width=0.75\textwidth]{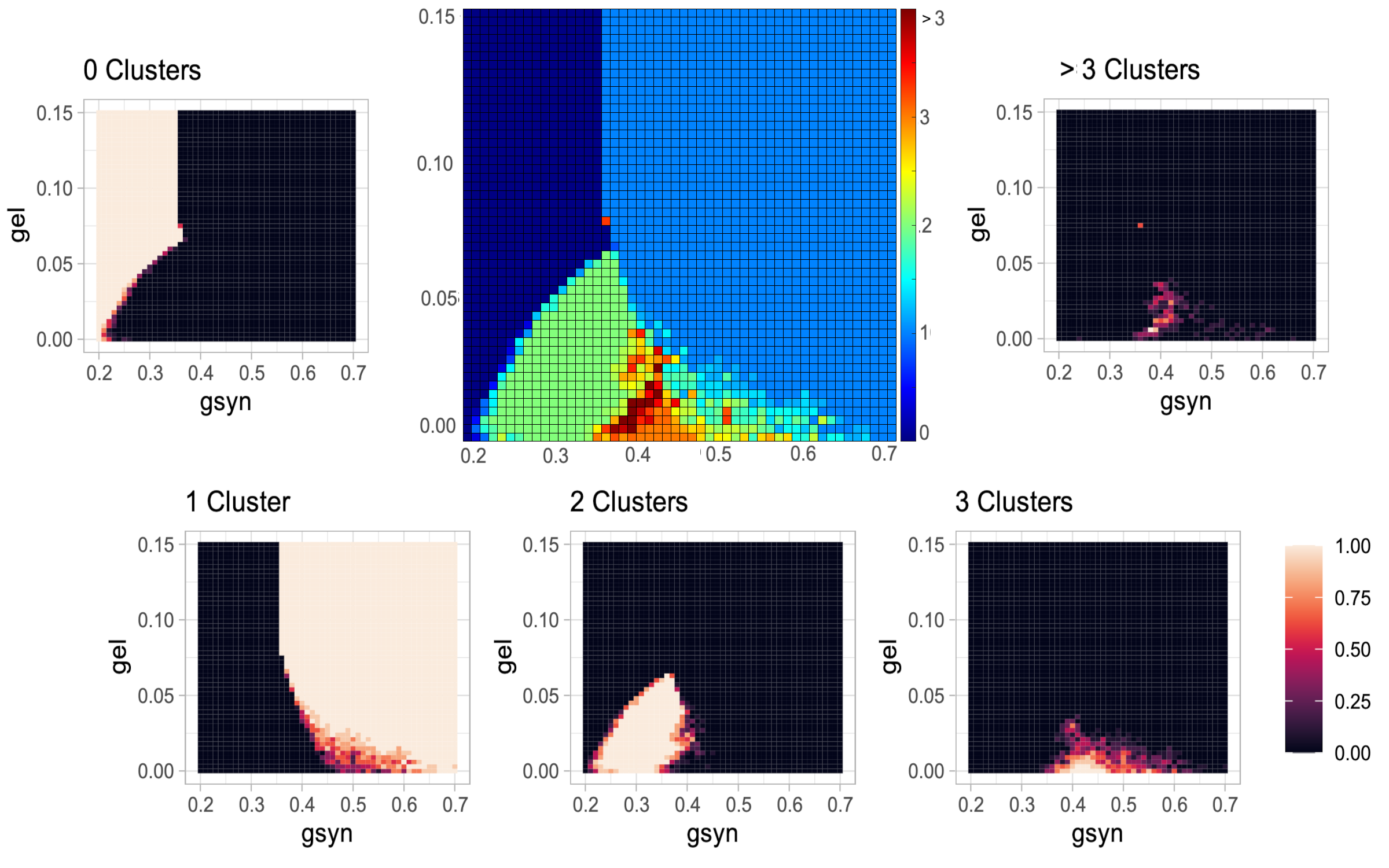}
\end{center}
\caption{{{\bf Likelihood of different synch-clustering behaviors in a region of the $(g_{syn},g_{el})$ parameter plane} for the case of all-to-all by gap junction connectivity. The central panel shows the average number of clusters for each parameter pair, over $K=11$ runs (with color coding specified in the color bar). The actual distribution of synch-cluster behaviors is detailed in the five surrounding panels, showing frequency plots for each of five possible behaviors (0,1,2,3, and $>3$ synch-clusters). Each frequency plot illustrates, for every point $(g_{syn},g_{el}) \in [0.2,0.7] \times [0,0.15]$, the fraction of the $K=11$ random runs under the specified parameters that exhibited the given behavior out of the five possibilities as labeled).}}
\label{bif1}
\end{figure}

These first results suggest that is would be useful to have a more comprehensive representation of the system's dependence on both conductance parameters $g_{syn}$ and $g_{el}$ simultaneously. Since in most cases the system does not have a unique asymptotic attractor, we will be using stochastic bifurcation diagrams, as introduced in previous work~\cite{radulescu2015nonlinear}, to represent likelihood of specific behaviors as we navigate the $(g_{syn},g_{el})$ parameter plane. For our simulations, we considered the parameter region $[0.2,0.7] \times [0,0.15]$, and formed a grid by using partitions with 50 steps in each parameter direction. For each of the $50 \times 50$ parameter pairs in the partition, we tested for the possible behaviors based on a sample of 11 runs, starting at random initial conditions. Refining the grid and increasing the sample size would of course refine our results, but would concomitantly come with increased computational cost. For the current purposes, the resolution parameters we used proved sufficient to illustrate our points. The results are shown in Figure~\ref{bif1}.

\begin{figure}[h!]
\begin{center}
\includegraphics[width=\textwidth]{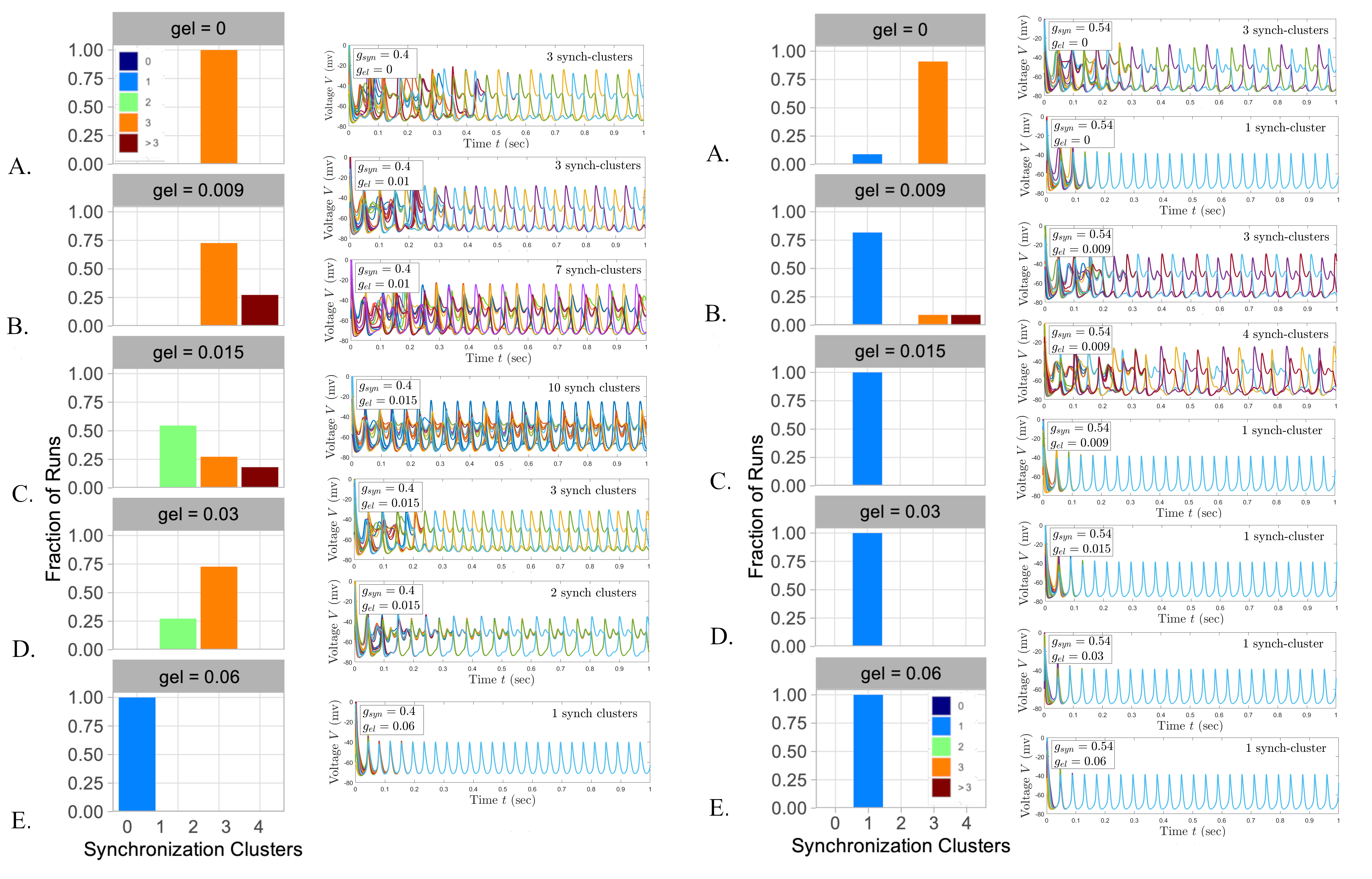}
\end{center}
\caption{{{\bf Changes in clustering behavior when $g_{el}$ increases}, for fixed $g_{syn}=0.4$ (panels A1-E1) and for $g_{syn}=0.54$ (panels A2-E2). The left panels show the fraction of each behavior based on a sample of $K=11$ runs with random initial conditions, and the right panels illustrate examples of those behaviors in $(V,t)$ coordinate axes. From top to bottom, the value of $g_{el}$ increases, as shown in the legends.}}
\label{vertical_slices}
\end{figure}

The figure confirms that, as suspected, the situation changes dramatically with increasing $g_{syn}$ and $g_{el}$. In particular, as suggested in Figure~\ref{horizontal_slices}, the system initially gains the ability to form three or more synch-clusters (within a small window of intermediate values of $g_{syn}$) when small values of $g_{el}$ are introduced. This is important, since the original Rinzel model, operated under the same networks size, did not exhibit this feature. However, as $g_{el}$ is increased past $~\sim 0.04$, the system loses this ability, and can form up to two synch-clusters only up to $g_{el} \sim 0.06$. For this range of $g_{el}$, the two synch-cluster behavior emerges within an intermediate $g_{syn}$ regime, and increasing $g_{syn}$ past this window when keeping $g_{el}$ fixed eventually transitions the system into full synchronization. If $g_{el}$ is increased even more (past $g_{el}\sim 0.06$), the system completely loses the ability to form multiple synch-clusters: provided $g_{syn}$ is high enough to produce oscillations, all variables are fully synchronized.

From a different angle, one can observe what happens when increasing $g_{el}$ for a fixed background synaptic weight $g_{syn}$. Clearly, the effects depend crucially on the fixed value of $g_{syn}$. For smaller values of $g_{syn}$, the two-cluster synchronization at low gap junctional conductance levels is eventually suppressed to damped oscillations when increasing $g_{el}$. Up to a critical value of $g_{syn} \sim 0.37$, increasing $g_{syn}$ requires a larger $g_{el}$ to accomplish this transition. Once the critical value of $g_{syn}$ is crossed, increasing $g_{el}$ has a qualitatively different effect. At first, the de-synchronization may increase, with the system showing three or more clusters (which is counter-intuitive, and conflicts with the general expectation that ``gap junctions increase synchronization''). However, when $g_{el}$ is increased even more, the system eventually becomes fully synchronized (which is in line with what one would generally expect and with predictions from other modeling studies).

As a simplification, one can say that increasing the gap junctional strength $g_{el}$ does \emph{eventually} synchronize the system (whether by damping the oscillations to a stable equilibrium, or by synchronizing persistent oscillations to one phase coherent cluster). However, this is a coarse description of a more complex picture, and the dependence of synchronization on $g_{el}$ depends on other circumstances, explored in the next sections.
    
\subsection{Other gap-cluster structures}
\subsubsection{Bifurcations and frequency plots}

\begin{landscape}
\begin{figure}[h!]
\begin{center}
\includegraphics[width=0.6\textwidth]{figures/bifurcations1.png}
\quad \quad
\includegraphics[width=0.6\textwidth]{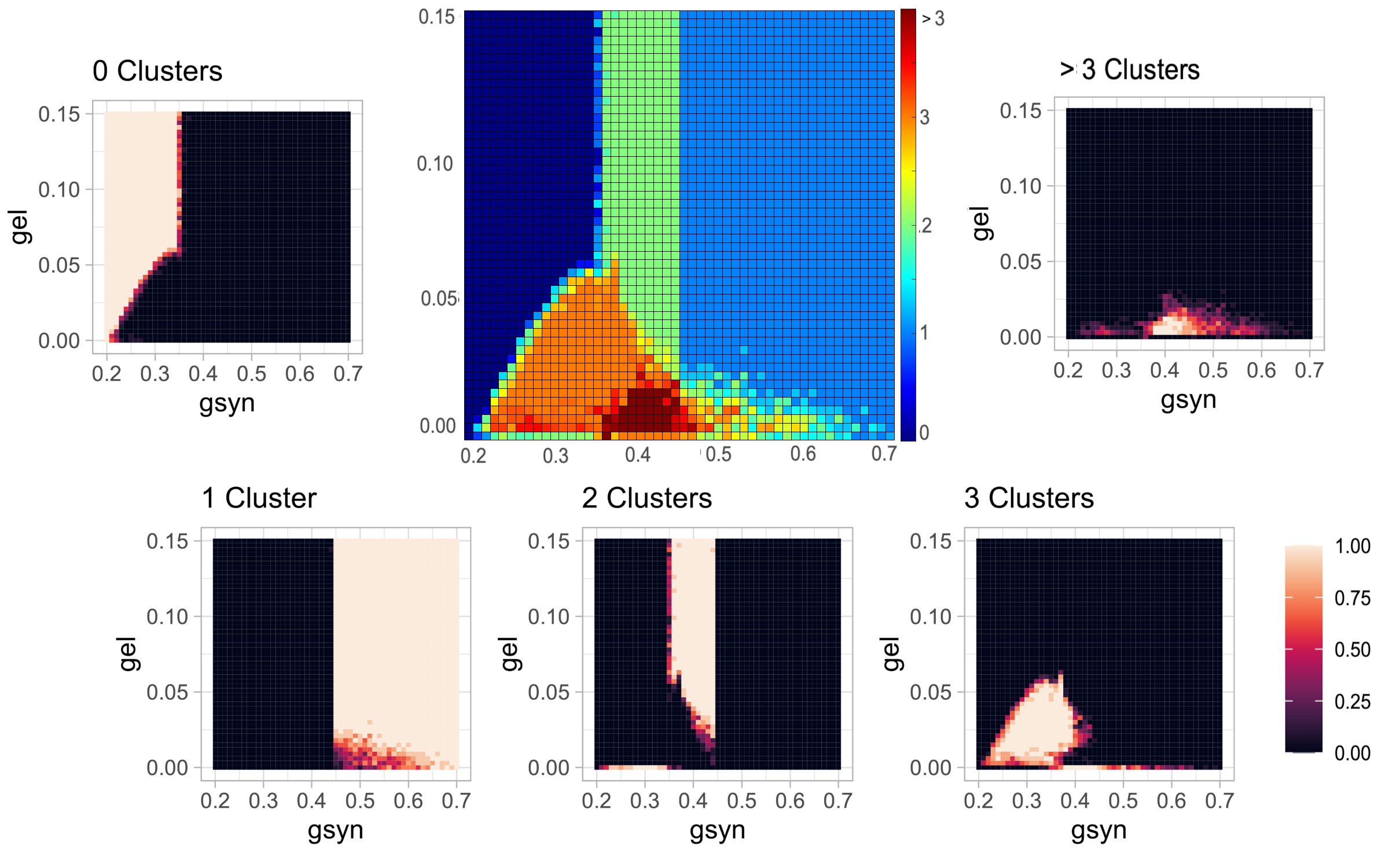}

\vspace{1cm}
\includegraphics[width=0.6\textwidth]{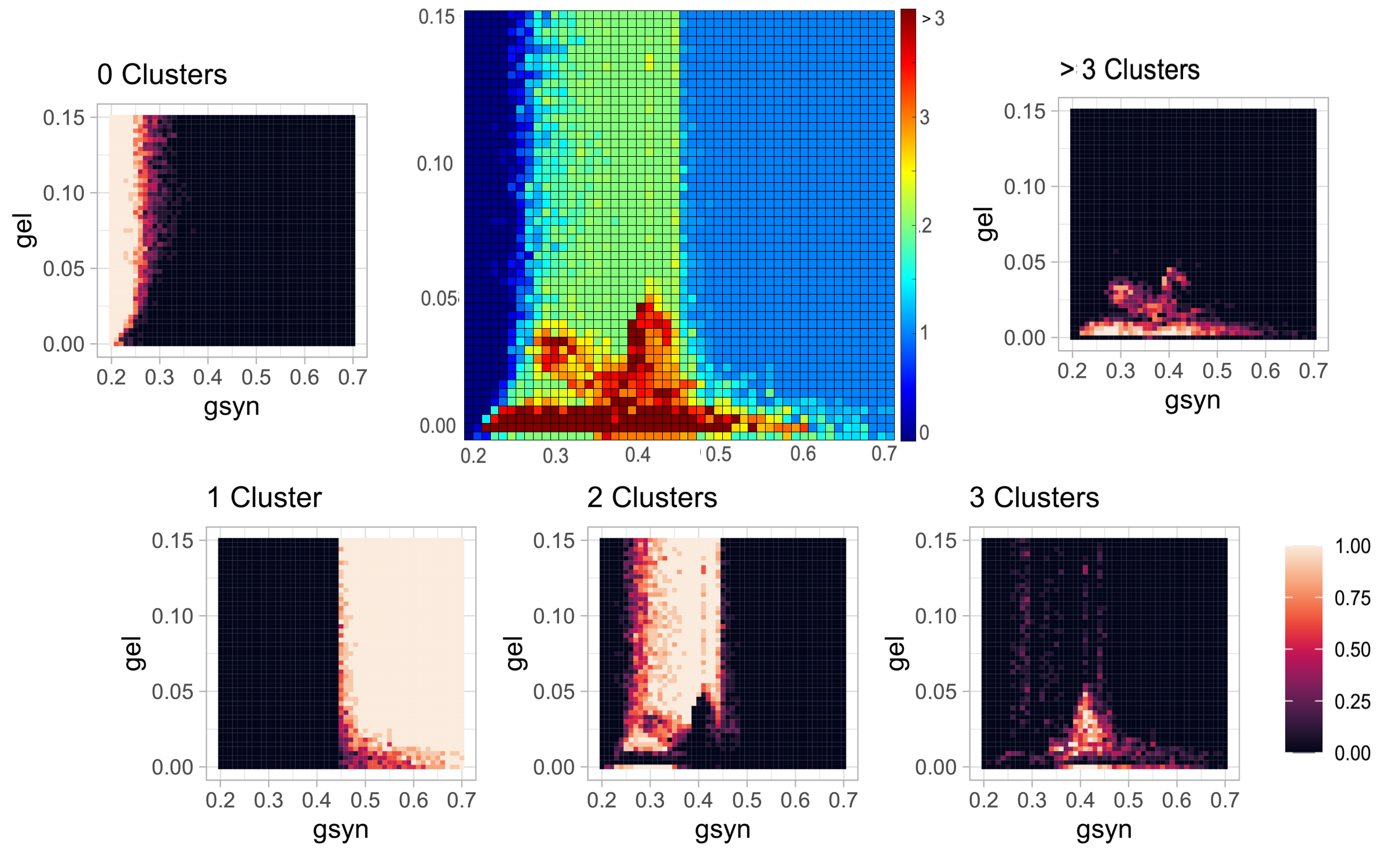}
\quad \quad
\includegraphics[width=0.6\textwidth]{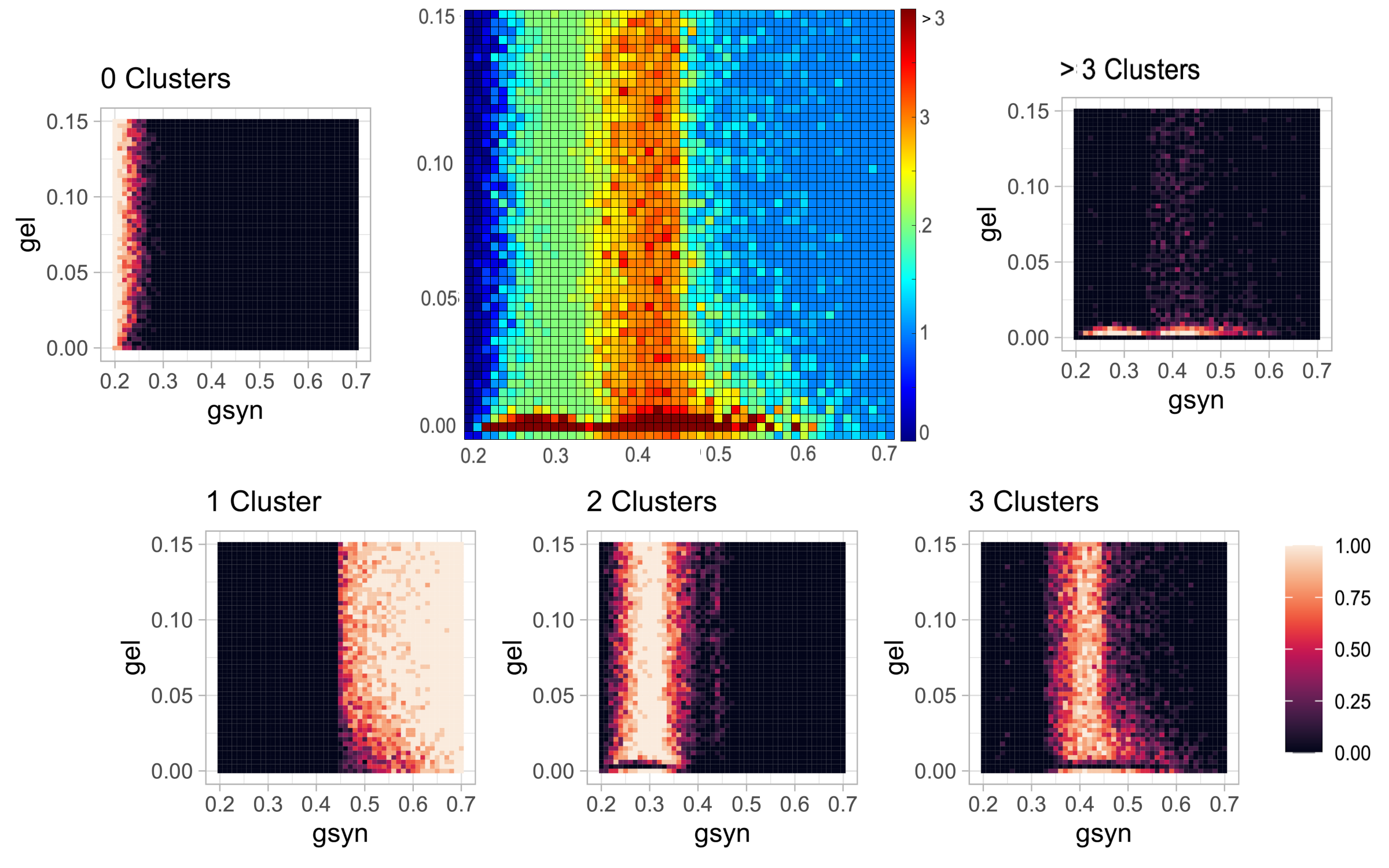}
\end{center}
\caption{\it{\small {\bf Likelihood of different synch-clustering behaviors in a region of the $(g_{syn},g_{el})$ parameter plane} for the three cases of gap junction connectivity architecture described in the Methods section. The panels represent the same concepts as in Figure~\ref{bif1} for the all-to-all case, with the same color coding.}}
\label{bif2}
\end{figure}
\end{landscape}

The previous section considered the case of all-to-all connectivity through gap junctions (so that each neuron pair is connected through both chemical and electrical synapses). This is, of course, a simplification, and does not correspond to the reality from empirical studies, as described in the Introduction, in which gap junctions organize in connected patches, which we call gap-clusters. The story becomes even more interesting when considering realistic distributions of electrical connections. In our exploration, we considered the arrangements described in Figure~\ref{networks}, and observed how to transitions in the $(g_{syn},g_{el})$ parameter plane change when the electrical connectivity matrix changes. We illustrate all four examples in Figure~\ref{bif2}.

In Figure~\ref{networks}b, the all-to-all gap connectivity is only slightly disturbed to a small gap-cluster $C_1$ (two nodes), and a large one $C_2$ (17 nodes). The modifications triggered in phase synchronization are illustrated in Figure~\ref{bif2}B. While the part of the plane and transitions remains unchanged, let's highlight the few very important changes. One difference is the presence of a large region where three synch-clusters are prevalent (orange region, as opposed to the two synch-clusters green region in Figure~\ref{bif2}a), with a small kernel showing even higher de-synchronization (dark red). This already shows that a small change to the all-to-all electrical connectivity already gives the system access to richer behavior. This being said, for small and large values of $g_{syn}$, increasing $g_{el}$ accomplishes the same effect as before: eliminates the synch-clusters by either damping oscillations (dark blue region) or by accomplishing full synchronization (light blue region). However, for intermediate values of $g_{syn}$ approximately between 0.35 and 0.45, increasing $g_{el}$ past a threshold point switches the oscillations to a two synch-cluster regime, which is robust to further increasing $g_{el}$ (the green straight central column). It is interesting to notice that this scenario suggests that there are limitations to the levels of synchronization that can be accomplished by simply increasing electrical connections strength: for certain values of $g_{syn}$, synchronization cannot be tightened beyond two synch-clusters, no matter how large $g_{el}$ might be.

We found a modified scenario when we used the architecture described in Figure~\ref{networks}c, which balances a little more the size of the two clusters: $C_1$ (four cells) and $C_2$ (14 cells). Indeed, the parameter plane in Figure~\ref{bif2}c shows a more extensive interval for $g_{syn}$ (approximately 0.25 to 0.45) for which the two synch-cluster structure persists when increasing $g_{el}$. In addition, the well-contoured three synch-cluster orange region disappeared, and low values of $g_{el}$ show a higher synch-cluster behavior (dark red), transitioning more rapidly into two synch-clusters when $g_{el}$ is increased.

The situation changes more dramatically when considering instead the electrical connectivity scheme in Figure~\ref{networks}d, with three gap-clusters $C_1$ (four nodes), $C_2$ (three nodes), $C_3$ (three nodes) and $C_4$ (five nodes). Notice that the corresponding Figure~\ref{bif2}d shows a ``column'' with primarily two synch-clusters, similar to the one in the previous case, but which is initiated at lower values of $g_{syn}$, stretching over $g_{syn} \sim 0.22-0.38$ and persists as $g_{el}$ is increased indefinitely. However, for this case, this is followed by a second column with primarily three synch-clusters for $g_{syn} \sim 0.38-0.45$. In addition, notice that the transition from the region of high likelihood of three synch-clusters to full synchronization still occurs when $g_{syn}$ is larger than 0.45, but is it a lot more gradual. There is a relatively broad interval for $g_{syn}$ (larger for smaller $g_{el}$ and decreasing for higher $g_{el}$) for which the three synch-cluster and one synch-cluster scenarios coexist.

\subsubsection{Navigating between architectures}

\begin{figure}[h!]
\begin{center}
\includegraphics[width=0.65\textwidth]{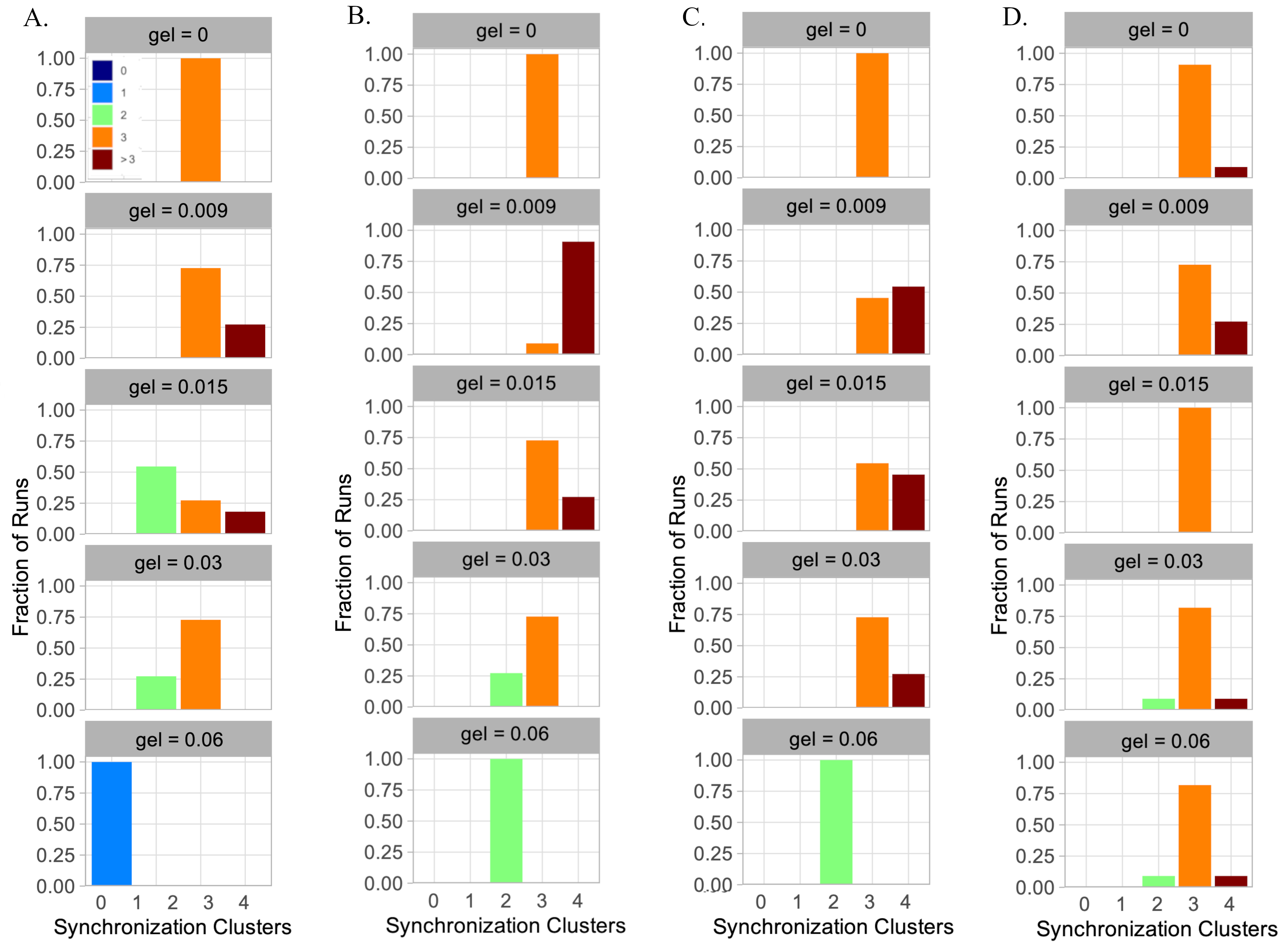}
\end{center}
\caption{{{\bf Changes in clustering behavior when changing the gap junction connectivity pattern the gap-cluster structure,} for a lower inhibitory conductance $g_{syn}=0.4$. Each column represents one connectivity scheme, in the order presented in Figure~\ref{networks}; the likelihood of synchronization patterns is shown for different values of $g_{el}$, for comparison.}}
\label{slice1}
\end{figure}

All of this clearly shows that the number, size and placement of gap-clusters are strong determinants of the system's ultimate dynamics, and to the formation of different synchronization patterns. To point out the extent to which these differences supplement the conductance parameters in governing the system's behavior, we show in Figures~\ref{slice1} and~\ref{slice2} direct comparisons between the synch-clustering patterns in the four connectivity schemes considered in Figure~\ref{networks} and~\ref{bif2}, for two fixed values of $g_{syn}$. In Figure~\ref{slice1}, we fix $g_{syn}=0.4$, and show how the synch-clusters reorganizes as $g_{el}$ is increased, illustrating one particular gap-clustering scheme in each column. There are dramatic differences in dynamic behavior between gap-cluster architectures. For our first connectivity scheme (all-to-all gap connectivity), increasing the strength of gap junctions gradually transitions the system from a regime of three synch clusters to full synchronization. For the same interval of $g_{el}$, the second and third connectivity schemes (involving two gap-clusters) can only push the system from three synch-clusters to two synch-clusters. Finally, for our last connectivity architecture (involving four gap-clusters), the three synch-cluster behavior remains predominant in the system for all values of $g_{el}$.

In contrast, the same analysis for a section corresponding to a higher $g_{syn} = 0.54$ shows only subtle differences in behavior between gap junction distribution schemes(Figure~\ref{slice1}). This reiterates the importance of the underlying context of the existing chemical synapses in determining how much changes in gap junction distribution and strength affect the phase synchronization of the network.

\section{Discussion}
In this paper, we investigated how the number, strength and architecture of electrical connections work together with the background inhibitory synapses between inter-neurons to orchestrate synchronization rhythms in the TRN. While many studies have shown that, broadly speaking, the effect of increasing gap junction connectivity pushes the system towards increased synchronization, our exploration shows that the meaning and degree of this synchronization strengthening depends crucially on context. To establish finer aspects of this dependence in small networks ($N=20$ neurons in our explorations), we did not stop at analyzing the effects of fewer versus denser gap junctional connections, but we also considered broad ranges of both chemical and electrical synaptic conductances, as well as different gap junction architectures (with neurons connected by gap junctions organized different numbers of clusters of different sizes).

When fixing the network connectivity scheme, we found that lower versus higher background inhibition dictates how the system will respond to increasing electrical conductance. Broadly speaking, the system evolves towards increased phase coherence, in the sense that, when $g_{el}$ is increased enough, the number of (spiking or sub-threshold) synchronization clusters eventually gets smaller. However, the path to increased phase coherence depends on the value of $g_{syn}$. For example, for small enough values of $g_{syn}$, increasing $g_{el}$ always results in damped oscillations. In turn, for large enough $g_{syn}$, increasing $g_{el}$ always results in fully phase-coherent oscillations (although the transition may slowly evolve through in intermediate stages where different cluster patterns coexist). 

Unsurprisingly, the story is a lot more complex for intermediate values of $g_{syn}$. There are a few interesting, and rather unexpected effects of increasing $g_{el}$ in this case. First, notice that the path of increasing $g_{el}$ crosses through intermediate ranges in which the system briefly becomes more de-synchronized (showing four or more synch-clusters by the end of the temporal window observed in our simulations). Since these clusters were only observed numerically for a limited time, there exists, of course, the possibility that the de-synchronization is transient, and that an analytic approach, or a much longer observation would report a ``chimera-like'' scenario, with a qualitatively different asymptotic behavior. However, this doesn't change the fact that the 

\begin{figure}[h!]
\begin{center}
\includegraphics[width=0.65\textwidth]{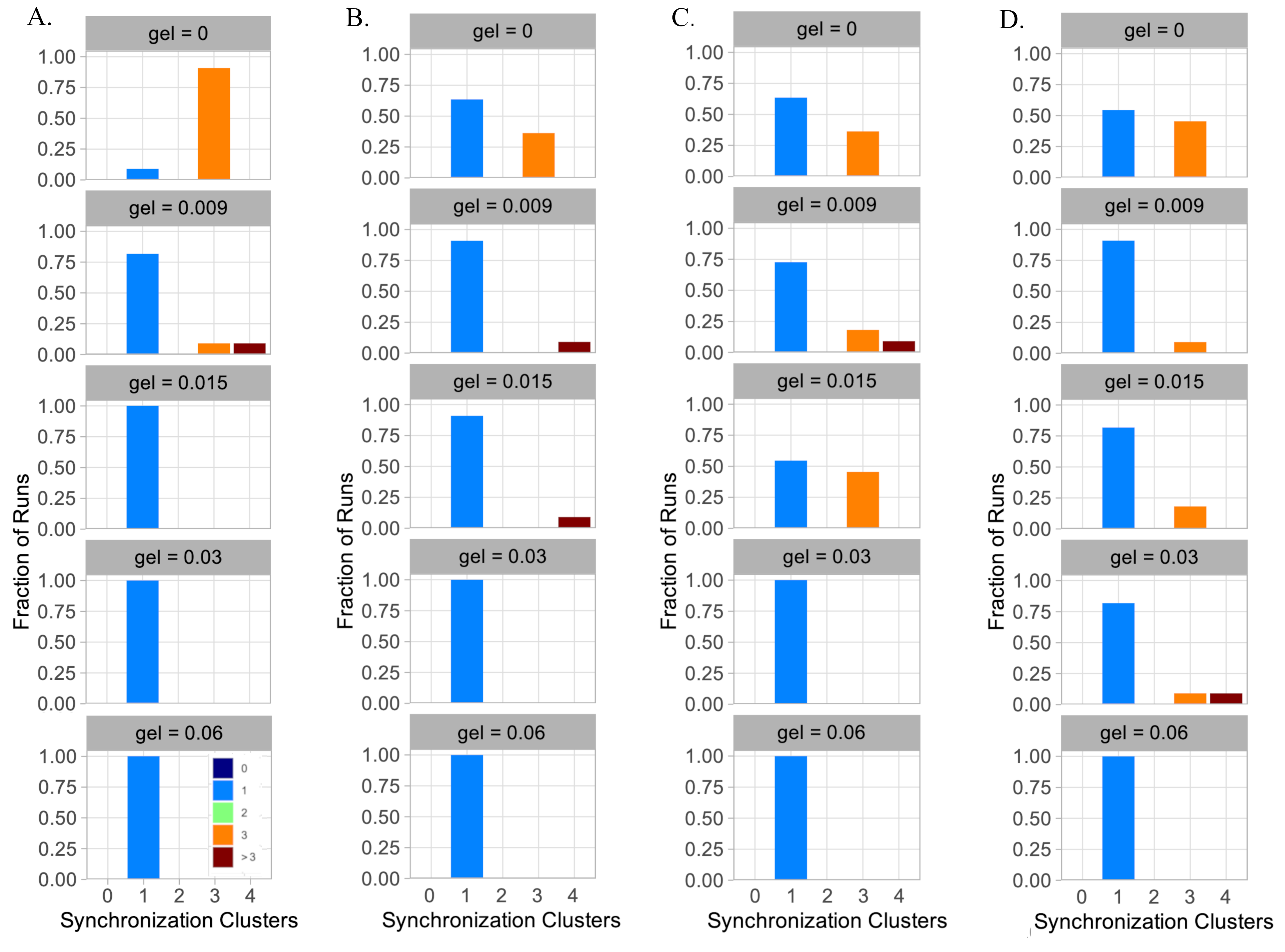}
\end{center}
\caption{{{\bf Changes in clustering behavior when changing the gap junction connectivity pattern the gap-cluster structure,} for a higher inhibitory conductance $g_{syn}=0.54$. Each column represents one connectivity scheme, in the order presented in Figure~\ref{networks}; the likelihood of synchronization patterns is shown for different values of $g_{el}$, for comparison.}}
\label{slice2}
\end{figure}

\noindent system showed this behavior only for a relatively slim range of small $g_{el}$ values, at least for the duration of the time window of the simulation, and that these multi-clusters did not persist for larger $g_{el}$.

A second interesting feature related to the gap-cluster architecture is that, for intermediate $g_{syn}$, the number of synch-clusters is tightly connected with the number of gap-clusters accessible to the system for large $g_{el}$. Indeed, the all-to-all connectivity scheme only allows two scenarios: damped oscillations, and full phase synchrony. The two architectures with two gap-clusters also promote the presence of two synch-clusters, and the four-cluster architecture allows for the possibility of two synch-clusters (for smaller values of $g_{syn})$, but also for three synch-clusters (for values of $g_{syn}$ a little larger).

Finally, we would like to comment on the colored ``vertical columns'' apparent in our $(g_{syn},g_{el})$ parameter planes, for the configurations involving more than one gap-cluster. These represent the fact that, for a certain $g_{syn}$ interval, increasing $g_{el}$ past a certain point is not going to further contribute to synchronization. This ``saturation'' property is interesting, especially since it appears with variation for all three connectivity patterns. For example, the green columns corresponding to networks \#2 and \#3 tell us that within the corresponding range of inhibitory strength, increasing electrical conductance will first de-synchronize the system, but eventually it will force it into a persistent two synch-cluster behavior, which will never evolve into full phase coherence. Notice that the range of $g_{syn}$ for which this behavior occurs is larger in the case of the networks with two more size-balanced gap-clusters ($C_1=4$ and $C_2=14$ versus $C_1=2$ and $C_2=19$). The situation changes again when looking at the column corresponding to network \#4, which contains both two and three synch-clusters (separated quite sharply for smaller versus larger values of $g_{syn}$).

Altogether, it is interesting to see, on one hand, the wild differences in behavior induced by changing the gap-cluster structure, and, on the other hand, the persistence of well-separated clustering behavior of the system past a certain value of $g_{el}$ (which is independent of connectivity scheme, and looks munch like the behavior in the original Golumb-Rinzel model). While our results are generally in line with recent literature on TRN synchronization, they shed a new light on the importance of considering existing questions in a complex systems context, so that the focus is on potential ensemble behaviors, rather than a single behavior of the model, and so that dependence on any one parameter can be considered in the context of a wider set of circumstances.

\section*{Appendix A}

\begin{figure}[h!]
\begin{center}
\includegraphics[width=0.9\textwidth]{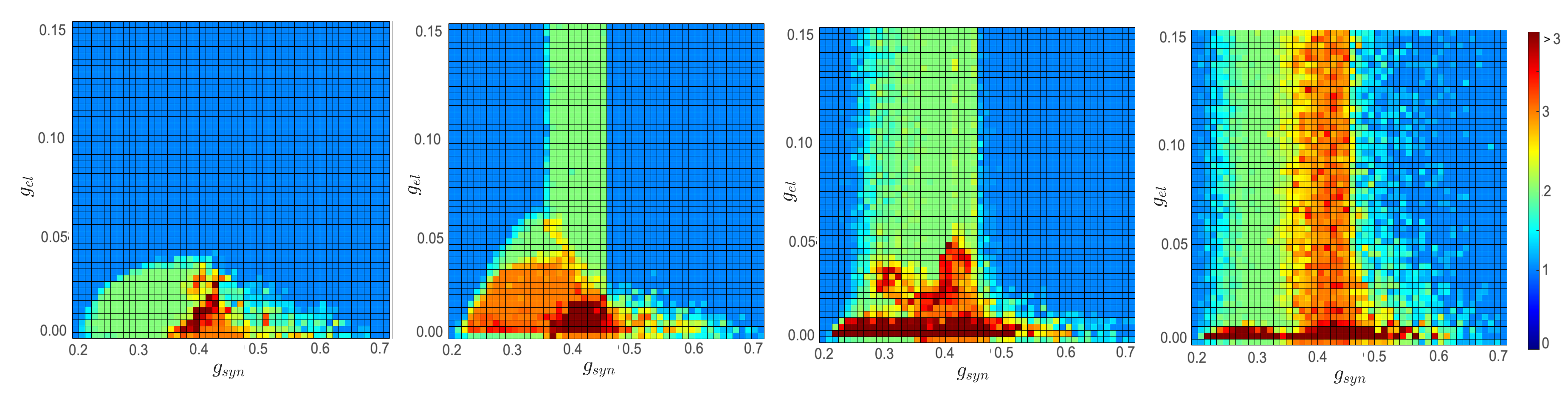}
\end{center}
\caption{{{\bf Changes in synchronization behavior in response to changing connectivity properties,} revised to disregard sub-threshold rhythms. This is a revision of the parameter planes in Figure~\ref{bif2}, so that the synch-clusters are counted according to a different scheme, where neurons with subtheshold rhytms are all counted as ``one synch-cluster'' (whether they perform damped oscillations, or sustained sub-threshold oscillations.}}
\label{slice2}
\end{figure}

\bibliographystyle{plain}
\bibliography{references}

\begin{thebibliography}{10}

\bibitem{connors2004electrical}
Barry~W Connors and Michael~A Long.
\newblock Electrical synapses in the mammalian brain.
\newblock {\em Annu. Rev. Neurosci.}, 27:393--418, 2004.

\bibitem{contreras1997spatiotemporal}
Diego Contreras, Alain Destexhe, Terrence~J Sejnowski, and Mircea Steriade.
\newblock Spatiotemporal patterns of spindle oscillations in cortex and
  thalamus.
\newblock {\em Journal of Neuroscience}, 17(3):1179--1196, 1997.

\bibitem{crabtree2018functional}
John~W Crabtree.
\newblock Functional diversity of thalamic reticular subnetworks.
\newblock {\em Frontiers in systems neuroscience}, 12:41, 2018.

\bibitem{golomb1993dynamics}
David Golomb and John Rinzel.
\newblock Dynamics of globally coupled inhibitory neurons with heterogeneity.
\newblock {\em Physical review E}, 48(6):4810, 1993.

\bibitem{golomb1994clustering}
David Golomb and John Rinzel.
\newblock Clustering in globally coupled inhibitory neurons.
\newblock {\em Physica D: Nonlinear Phenomena}, 72(3):259--282, 1994.

\bibitem{hjorth2009gap}
Johannes Hjorth, Kim~T Blackwell, and Jeanette~Hellgren Kotaleski.
\newblock Gap junctions between striatal fast-spiking interneurons regulate
  spiking activity and synchronization as a function of cortical activity.
\newblock {\em Journal of Neuroscience}, 29(16):5276--5286, 2009.

\bibitem{kopell2004chemical}
Nancy Kopell and Bard Ermentrout.
\newblock Chemical and electrical synapses perform complementary roles in the
  synchronization of interneuronal networks.
\newblock {\em Proceedings of the National Academy of Sciences},
  101(43):15482--15487, 2004.

\bibitem{krosigk1993cellular}
Marcus Krosigk~von, Thierry Bal, and David~A McCormick.
\newblock Cellular mechanisms of a synchronized oscillation in the thalamus.
\newblock {\em Science}, 261(5119):361--364, 1993.

\bibitem{lee2014two}
Seung-Chan Lee, Saundra~L Patrick, Kristen~A Richardson, and Barry~W Connors.
\newblock Two functionally distinct networks of gap junction-coupled inhibitory
  neurons in the thalamic reticular nucleus.
\newblock {\em Journal of Neuroscience}, 34(39):13170--13182, 2014.

\bibitem{long2004small}
Michael~A Long, Carole~E Landisman, and Barry~W Connors.
\newblock Small clusters of electrically coupled neurons generate synchronous
  rhythms in the thalamic reticular nucleus.
\newblock {\em Journal of Neuroscience}, 24(2):341--349, 2004.

\bibitem{mccormick2001cellular}
David~A McCormick and Diego Contreras.
\newblock On the cellular and network bases of epileptic seizures.
\newblock {\em Annual review of physiology}, 63:815, 2001.

\bibitem{radulescu2015nonlinear}
Anca R\v{a}dulescu and Sergio Verduzco-Flores.
\newblock Nonlinear network dynamics under perturbations of the underlying
  graph.
\newblock {\em Chaos: An Interdisciplinary Journal of Nonlinear Science},
  25(1):013116, 2015.

\bibitem{simon2005gap}
Anna Simon, Szabolcs Ol{\'a}h, G{\'a}bor Moln{\'a}r, J{\'a}nos Szabadics, and
  G{\'a}bor Tam{\'a}s.
\newblock Gap-junctional coupling between neurogliaform cells and various
  interneuron types in the neocortex.
\newblock {\em Journal of Neuroscience}, 25(27):6278--6285, 2005.

\bibitem{slaght2002activity}
Se{\'a}n~J Slaght, Nathalie Leresche, Jean-Michel Deniau, Vincenzo Crunelli,
  and St{\'e}phane Charpier.
\newblock Activity of thalamic reticular neurons during spontaneous genetically
  determined spike and wave discharges.
\newblock {\em Journal of Neuroscience}, 22(6):2323--2334, 2002.

\bibitem{steriade1993thalamocortical}
Mircea Steriade, David~A McCormick, and Terrence~J Sejnowski.
\newblock Thalamocortical oscillations in the sleeping and aroused brain.
\newblock {\em Science}, 262(5134):679--685, 1993.

\bibitem{tiesinga2008regulation}
Paul Tiesinga, Jean-Marc Fellous, and Terrence~J Sejnowski.
\newblock Regulation of spike timing in visual cortical circuits.
\newblock {\em Nature reviews neuroscience}, 9(2):97--107, 2008.

\bibitem{vervaeke2010rapid}
Koen Vervaeke, Andrea L{\H{o}}rincz, Padraig Gleeson, Matteo Farinella, Zoltan
  Nusser, and R~Angus Silver.
\newblock Rapid desynchronization of an electrically coupled interneuron
  network with sparse excitatory synaptic input.
\newblock {\em Neuron}, 67(3):435--451, 2010.

\end{thebibliography}

\end{document}